%% file: main.tex
\documentclass{article}
\usepackage{spconf,amsmath,graphicx,bm}

\usepackage{amssymb}
\usepackage{booktabs}
\usepackage[pagebackref,breaklinks,colorlinks]{hyperref}
\usepackage[ruled,linesnumbered]{algorithm2e}
\usepackage[sort&compress,numbers,comma]{natbib}

\usepackage{xcolor}

\def \eg {\emph{e.g.}}
\def \ie {\emph{i.e.}}

\title{Confidence-based Event-centric Online Video Question Answering on a Newly Constructed ATBS Dataset}

\name{Weikai Kong$^{\star}$$^{\ddagger}$, Shuhong Ye$^{\star}$$^{\ddagger}$,  Chenglin Yao$^{\star}$, Jianfeng Ren$^{\star\dagger}$ \thanks{$^{\ddagger}$ The authors contributed equally.} 
\thanks{This work was supported in part by the National Natural Science Foundation of China under Grant 72071116, and in part by the Ningbo Municipal Bureau Science and Technology under Grants 2019B10026 and 2022Z173. Corresponding author: jianfeng.ren@nottingham.edu.cn.}}
\address{${}^{\star}$School of Computer Science, University of Nottingham Ningbo China\\
$^{\dagger}$Nottingham Ningbo China Beacons of Excellence Research and Innovation Institute, \\
University of Nottingham Ningbo China, China}

\begin{document}
%
\maketitle
\begin{abstract}
Deep neural networks facilitate video question answering (VideoQA), but the real-world applications on video streams such as CCTV and live cast place higher demands on the solver. To address the challenges of VideoQA on long videos of unknown length, we define a new set of problems called Online Open-ended Video Question Answering (O$^{2}$VQA). It requires an online state-updating mechanism for the solver to decide if the collected information is sufficient to conclude an answer. We then propose a Confidence-based Event-centric Online Video Question Answering (CEO-VQA) model to solve this problem. Furthermore, a dataset called Answer Target in Background Stream (ATBS) is constructed to evaluate this newly developed online VideoQA application. Compared to the baseline VideoQA method that watches the whole video, the experimental results show that the proposed method achieves a significant performance gain. 
\end{abstract}
\section{Introduction}

Video Question Answering (VideoQA) 
is beneficial to many real-life applications, \eg, intelligent conversation \cite{wang2021pairwise_application}, multimedia recommendation \cite{ theodoridis2013music_icassp}, and web video summarization \cite{wang2012event_video_summarization}. In open-ended VideoQA tasks, a video and a related question are provided and the model needs to provide the right answer. 
Recent methods \cite{li2019learnable_coattention,gao2019structured_coattention,zha2019spatiotemporal_coattention,jin2019multi_selfattention} often use the attention mechanism to model the interaction between videos and questions. 

\begin{figure*}[htpb]
\centering
\includegraphics[width=0.88\textwidth]{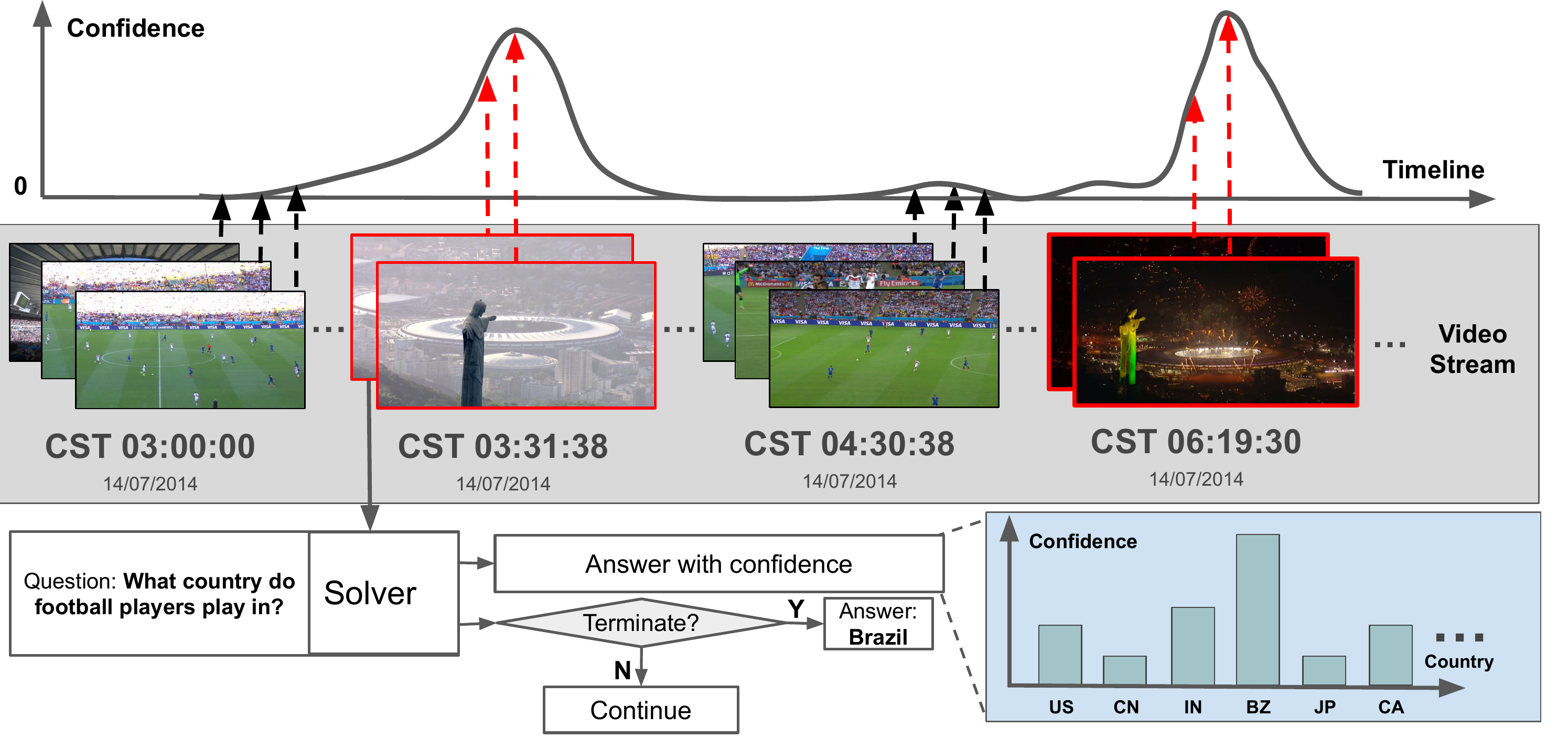}
  \caption{Illustration of Online Open-ended
Video Question Answering (O$^2$VQA) task. 
For each frame, the solver derives the confidence score for each of the feasible candidate answers and decides if the evidence is sufficient to give a confident answer.}
  \label{fig:1}
\end{figure*}

Most existing models focus on answering questions given short videos of a fixed length \cite{le2022hierarchical_HCRN_relation_model_benchmark,xiao2022video_HQGA_SOTA,fan2019heterogeneous_HME_mem_att_memory_model_benchmark,li2019learnable_coattention,gao2019structured_coattention,zha2019spatiotemporal_coattention,jin2019multi_selfattention,gao2018motion_CoMem_mem_SOTA}. But in real-world scenarios, the video may be streamed from an unpredictably lengthy CCTV or live cast. 
Online videoQA posts three challenges: 1) oversized information 
that dilutes the key features, 2) disturbance events that may confuse the solver, and 3) lack of a mechanism 
to confirm that the collected information is enough to answer the question.
We model it as \textbf{O}nline \textbf{O}pen-ended \textbf{V}ideo \textbf{Q}uestion \textbf{A}nswering (O$^2$VQA) problem, as shown in Fig~\ref{fig:1}. 
Given a video of an unpredictable length and a relevant question, the solver shall online process the video, terminate when enough information is collected, and answer the question with a human-readable sentence. 

To tackle these problems, we propose a Confidence-based Event-centric Online Video Question Answering (CEO-VQA) model for O$^2$VQA tasks, which consists of two main modules: 1) Confidence-based Target Event Locator to locate the target event that is most relevant to answer the question; 2) Target Event Question Answering model to answer the question using the visual and language clues. 
Specifically, a BridgeFormer \cite{ge2022bridging_video_text_retrieval} 
is utilized to derive the similarity between the linguistic features and the visual features. The confidence score is then updated based on these similarities.  
The Confidence-based Target Event Locator then filters the irrelevant events and locates the target events. 
Finally, the visual features and the language features are integrated into the Target Event Question Answering model and fed into an answer decoder to derive the correct answer. 

To facilitate the evaluation of this newly developed O$^2$VQA task, a large video dataset is constructed. We use \emph{Background + Target} approach to generate the new dataset,  \textbf{A}nswer \textbf{T}arget in \textbf{B}ackground \textbf{S}tream (ATBS), to simulate complex multi-event video streaming. 
Specifically, we insert each of the 10k \emph{Target} video clips of the MSRVTT dataset \cite{xu2016msr_dataset}  into one of the 10,464 \emph{Background} videos of the DiDeMo dataset \cite{anne2017localizing_dataset_ddm} and assign the question-answer pairs of the \emph{Target} videos as the annotations. As a result, the constructed dataset contains 10k videos in total with 244k annotated QA pairs. As shown in the experiments, the proposed dataset is more challenging than existing VideoQA datasets \cite{xu2017video_att_MSVD-QA_MSRVTT-QA_end-to-end_attention_model, xu2016msr_dataset}. 

Our contributions are three-fold: 1) To advance the research in VideoQA, we introduce a novel and challenging VideoQA task, Online Open-ended Video Question Answering (O$^2$VQA); 2) We construct a large ATBS dataset to facilitate the research on O$^2$VQA tasks; 3) We propose a CEO-VQA model to more effectively solve the O$^2$VQA problems. 
\section{Proposed Method}
\subsection{Overview of Proposed Method}

To tackle the challenges of O$^2$VQA, we propose a Confidence-based Event-centric Online Video Question Answering (CEO-VQA) model. The main procedures are outlined in Algo. \ref{algo:1}. 
The proposed CEO-VQA consists of two main modules: 
1) Confidence-based Target Event Locator. Given an online video stream and a question, the confidence to answer is derived by using the BridgeFormer \cite{ge2022bridging_video_text_retrieval} to evaluate the attentional information between the text features extracted from the question and the visual features extracted from the video stream. The Confidence-based Target Event Locator then filters out irrelevant events and locates the target event. 
To reduce the time complexity, we bi-directionally traverse the video in a Fibonacci way. 
2) Target Event Question Answering. 
After locating the target event, a TimeSformer \cite{bertasius2021space_timesformer} is utilized to extract the visual features from the video key frames, and a BERT-base model \cite{devlin-etal-2019-bert, liu2022cross} is utilized to extract the linguistic features. The visual and linguistic features are then concatenated and fed into a multi-modal encoder to perform cross-modal learning. Finally, a reliable answer is generated by the answer decoder \cite{fan2019heterogeneous_HME_mem_att_memory_model_benchmark, le2022hierarchical_HCRN_relation_model_benchmark}.



\subsection{Confidence-based Target Event Locator}
Unlike existing VideoQA models \cite{li2022align_alpro, le2022hierarchical_HCRN_relation_model_benchmark,fan2019heterogeneous_HME_mem_att_memory_model_benchmark,xu2017video_att_MSVD-QA_MSRVTT-QA_end-to-end_attention_model} that encode the whole video, online video streaming such as live sports needs to answer a question with low latency given the streaming video. In addition, the comprehension of visual information is often disturbed by irrelevant events, especially in lengthy videos. We hence propose a Confidence-based Target Event Locator (CTEL) to filter unrelated events and find the target event. Compared to event localization 
\cite{zhang2019exploiting_locator,zhang2020learning_locator}, the question provides a much weaker language supervision than the given text description in event localization. 

The confidence in answering the question is derived based on the attentional information between the current frame and the given question using BridgeFormer \cite{ge2022bridging_video_text_retrieval}. Many deep neural networks have been developed for visual recognition \cite{yao2021rppg, song2023solving,song2023solving2}. In this paper, the visual features are extracted using a 12-layer vision transformer \cite{VIT,chen2022attention, he2023hierarchical}. It splits each frame into $N_{u}$ patches and linearly projects the patches to a sequence of visual tokens. A classification token [CLS] is added to the visual tokens. 
After adding learnable positional embeddings to the tokens, spatial self-attention is performed, and a sequence of visual features 
$\bm{U} = [\bm{u}_{cls}, \bm{u}_{1},..., \bm{u}_{N_{u}}]$ are generated. 
Similarly, a 12-layer text transformer, which adopts the architecture of DistilBERT \cite{sanh2019distilbert}, is used to encode the question. The question is tokenized into
$N_{p}$ 
tokens, concatenated with a [CLS] token, and added with learnable positional embeddings. Then self-attention is performed on the text tokens and a sequence of features
$\bm{P} = [\bm{p}_{cls}, \bm{p}_{1},..., \bm{p}_{N_{p}}]$ are generated. 
Finally, the [CLS] token of the visual feature 
$u_{cls}$ and 
question feature $p_{cls}$ are normalized and multiplied using dot product to obtain the confidence of the frame.



To remove disturbing events and locate the target event, 
a hysteresis thresholding scheme is developed, where $c_{min}$ represents the minimal expected confidence and  
$c_{max}$ represents the threshold for sufficient confidence. 
Once the confidence of the current frame is high enough, 
$c \geq c_{max}$, we bi-directionally traverse the video from the current frame in a Fibonacci way, until $c < c_{min}$. 
The traversed video frames $\mathcal{T}_{B}$ and $\mathcal{T}_{F}$ form the target event $\mathcal{T}$. The Fibonacci sampling is mainly to reduce the amount of computation. 

\input{sections/code}

\subsection{Target Event Question Answering }
After the target event $\mathcal{T}$ is located, an online Target Event Question Answering model is developed to generate the correct answer. 
The frames in the target event and the given question are encoded using a 12-layer TimeSformer \cite{bertasius2021space_timesformer} and a BERT-base model \cite{devlin-etal-2019-bert, liu2022cross} respectively to extract vision and language features. Each frame is patched and linearly projected to patch tokens. Then the tokens are added with learnable positional and temporal embeddings, and fed to the TimeSformer attention blocks to perform the divided spatial-temporal attention. Feature vectors of the encoded frames are average-pooled into a feature vector $\bm{V} = [\bm{v}_{cls}, \bm{v}_{1},..., \bm{v}_{N_{v}}]$, where $\bm{v}_{cls}$ is the [CLS] token used for classification. 
For the question encoding, the question is tokenized into a sequence with $N_{q}$ tokens. Then the self-attention is performed on the input question tokens, and a feature sequence $\bm{Q} = [\bm{q}_{cls}, \bm{q}_{1},..., \bm{q}_{N_{q}}]$ is derived as the output.

The derived text features $\bm{Q}$ and the visual features $\bm{V}$ are
concatenated into one sequence 
and fed into a multi-modal transformer \cite{li2022align_alpro} to perform self-attention across features without adding other embeddings. The output sequence is derived as $\bm{M} = [\bm{m}_{cls}, \bm{m}_{1},..., \bm{m}_{N_{q} +N_{v}}]$. Finally, its $\bm{m}_{cls}$ [CLS] tokens are fused using MLPs and an answer decoder \cite{fan2019heterogeneous_HME_mem_att_memory_model_benchmark, le2022hierarchical_HCRN_relation_model_benchmark} is adopted to derive the correct answer.



\section{Construction of ATBS Dataset}
\begin{table*}[ht]
  \centering
    \caption{Comparison with existing VideoQA datasets.}
    \begin{tabular}{@{}lc c c c c c}
    \toprule
                   Dataset & Task    & Video composition   & Total duration & Average duration & \# Videos & \# QA pairs \\
    \midrule
     MSRVTT-QA \cite{xu2017video_att_MSVD-QA_MSRVTT-QA_end-to-end_attention_model} 
                           & VideoQA & Target only         & 38.89h   & 14s   & 10,000   & 243,680     \\
     MSVD-QA \cite{xu2017video_att_MSVD-QA_MSRVTT-QA_end-to-end_attention_model} 
                           & VideoQA & Target only         & 5.47h   & 10s   & 1,970   & 50,505  \\
     TGIF-QA \cite{jang2017tgif_dataset} 
                           & VideoQA & Target only         & 47.27h   & 3s   & 56,720          & 103,919  \\
     Video-QA \cite{zeng2017leveraging_dataset} 
                           & VideoQA & Target only         & 226.25h   & 45s   & 18,100          & 174,775  \\                       
    ATBS & O$^2$VQA    & Background + Target & 175h & 63s & 10,000           & 243,680  \\
    \bottomrule
    \end{tabular}
  \label{tab:1}
  
\end{table*}


To evaluate the proposed method for the newly developed O$^{2}$VQA task, we construct an \textbf{A}nswer \textbf{T}arget in \textbf{B}ackground \textbf{S}tream (ATBS) dataset\footnote{Available at: \url{https://kenn-san.github.io/ATBS}}. 
For the O$^2$VQA task, the visual clue to correctly answer the question may appear multiple times at any time with unknown durations. To simulate the real-world scenarios, 
we design a \emph{Background + Target} approach, where the \emph{Background} is a relatively long background video to simulate an online video stream and the \emph{Target} is a target short video clip with the necessary information to answer the given question. The \emph{Target} video is inserted into the \emph{Background} at a random location to compose a video sample.

\noindent\textbf{Background Videos}: To simulate the real-world scenario where multiple events appear in a video stream, we choose to use the Distinct Describable Moments (DiDeMo) \cite{anne2017localizing_dataset_ddm} dataset as the source of background videos. It has multiple distinct events in each of the videos. It has 10,464 videos and an average video duration of around 48 seconds. 

\noindent\textbf{Target Videos}: Video clips from the MSRVTT dataset \cite{xu2016msr_dataset} are used as target videos, where each clip is a distinct event. It has 10k video clips and an average duration of about 14s.

\noindent\textbf{Dataset Generation:} 
Every video clip from the MSRVTT dataset \cite{xu2016msr_dataset} is uniquely paired with a background video randomly selected from the DiDeMo dataset \cite{anne2017localizing_dataset_ddm}. 
Both the target video clip frames and background video frames are resized to $224\times224$. Then the target video clip is inserted behind a randomly selected frame of the background video to generate the resulting video in our dataset, which simulates the random appearance of the target event in the video stream. 
Then, the question-answer pairs of the \emph{Target} video clips are chosen as the annotations for the generated videos. It is noted that pairing similar videos may lead to multiple ``correct" answers whereas the background video may contain distracting ``correct" answers. Although random pairing may merge a target video with an irrelevant background video, it assures a unique answer to the question. The final dataset contains 10k videos with a total of 244k question-answer pairs.

\begin{figure}[htbp]
\begin{minipage}[t]{0.55\linewidth}
    \includegraphics[width=\linewidth]{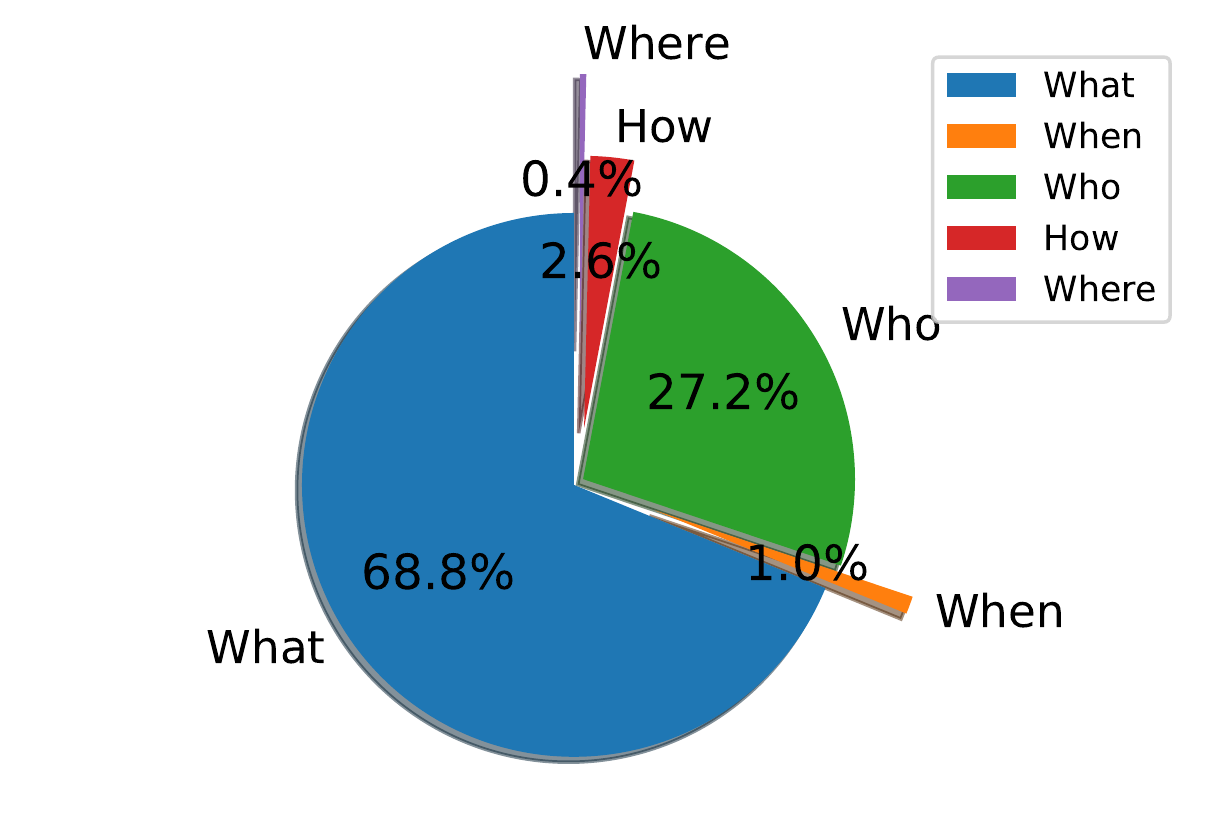}
    \label{f1}
\end{minipage}%
    \hfill
\begin{minipage}[t]{0.45\linewidth}
    \includegraphics[width=\linewidth]{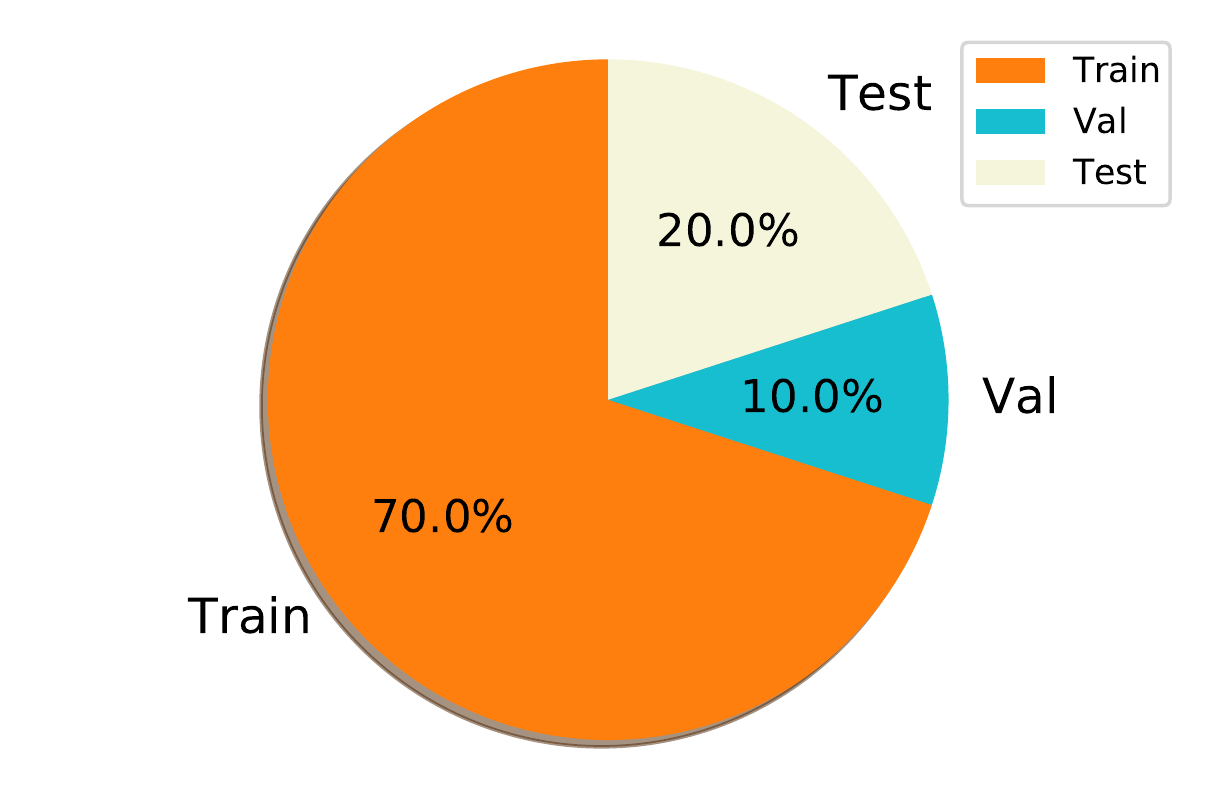}
    \label{f2}
\end{minipage} 
\vspace{-1.5em}
\caption{Distribution of the question types (left) and train-test split strategy (right) in the ATBS  dataset.}
\label{tab:3}

\end{figure}

The ATBS dataset is the first dataset for O$^2$VQA tasks. It is compared with existing VideoQA datasets, as summarized in Table \ref{tab:1}. 
The developed dataset has the longest average duration, the second longest total duration, and the largest number of QA pairs. 
There are 5 types of questions in the ATBS dataset. The distribution of the question types is summarized in Fig. \ref{tab:3}. 
The number of labels in the answer space is 1290.
For the dataset split, we select 7k videos as the training set, 1k videos for validation, and 2k videos for testing.


\section{Experimental Results}


\subsection{Experiment Settings}
As O$^2$VQA is a novel VideoQA task, there are no existing benchmark models. We hence adopt ALPRO \cite{li2022align_alpro}, the state-of-the-art VideoQA model, to the O$^2$VQA task. Compared with our method which only processes part of the video, ALPRO is allowed to encode the entire video. 
The video-text encoder in CTEL, BridgeFormer \cite{ge2022bridging_video_text_retrieval}, is initialized using the officially released weights\footnote{Available at:  \url{https://github.com/tencentarc/mcq}}.
We set the confidence thresholds empirically, \ie, $c_{max} =0.4$ and $c_{min} = 0.3$. 
The initial learning rate is set to 5e-5 and linearly decayed in subsequent epochs. The training batch size is 48. The AdamW optimizer is used with a weight decay of 1e-3. Experiments are performed on an NVIDIA A100 GPU. It takes about 3 GPU hours to train 20 epochs for both the proposed method and ALPRO. 

\subsection{Comparison Results}
We first compare the performance of the baseline model ALPRO~\cite{li2022align_alpro} on different datasets, to demonstrate the difficulty of our newly constructed ATBS dataset and the newly developed Q$^2$VQA task. 
As shown in Table \ref{tab:2}, the accuracy of ALPRO~\cite{li2022align_alpro} on the ATBS dataset is much lower than that on other datasets, which suggests that ATBS is more challenging as the useful visual cues are diluted in lengthy videos. 

%

    
    
\begin{table}[ht]
\setlength{\tabcolsep}{1mm}
  \centering
      \caption{Accuracy of ALPRO \cite{li2022align_alpro} on VideoQA datasets. Our ATBS dataset is much more challenging than others.}
    \begin{tabular}{@{}lcccccc}
    \toprule
    Dataset & what & who & how & where & when & overall \\
    \midrule
    MSVD-QA & 37.05 & 58.61 & 81.62 & 46.43 & 72.41 & 45.94    \\
    MSRVTT-QA & 36.05 & 52.24 & 85.67 & 42.80 & 78.88 & 42.12    \\
    \midrule
    \textbf{ATBS} & 19.16 & 33.58 & 77.27 & 25.00 & 54.55 & 24.80 \\
    \bottomrule
    \end{tabular}
    \label{tab:2}
    
\end{table}
%

Then, we compare the proposed model with ALPRO~\cite{li2022align_alpro} and \textbf{TargetEvent} on the ATBS dataset, where \textbf{TargetEvent} directly uses the ground-truth target event for VideoQA. As shown in Table~\ref{tab:3}, though our method only processes part of the videos, it boosts the overall performance of ALPRO~\cite{li2022align_alpro} from 24.8\% to 27.1\%.  The results demonstrate the effectiveness of the proposed method. Our method achieves superior results on the ``Who" and ``How" question types. Typically, it significantly outperforms the baseline model by 8.1\% on the ``Who" question type. This indicates that our method can locate the target event relevant to humans more precisely and give more accurate answers. For the ’How’, ’Where’ and ’When’ question types, our method achieves the same performance as \textbf{TargetEvent} that has direct access to the target event. Our method quickly locates the target event while ALPRO~\cite{li2022align_alpro} needs to answer the question using the whole video. As a result, our method runs faster than ALPRO, \eg, on a Tesla V100 GPU, our method needs 28.5s while ALPRO needs 63.6s when answering a question using one video.

\begin{table}[ht]
    \vspace{-.5em}
\setlength{\tabcolsep}{1mm}
  \centering
    \caption{Comparison between \textbf{TargetEvent}, ALPRO~\cite{li2022align_alpro} and the proposed \textbf{CEO-VQA} in top-1 accuracy (\%).}
  \resizebox{1\columnwidth}{!}{
    \begin{tabular}{@{}lcccccc}
    \toprule
    Method & what & who & how & where & when & overall \\
    \midrule
    TargetEvent & 24.38 & 43.25 & 79.55 & 25.0 & 54.55 & 31.1    \\
    ALPRO~\cite{li2022align_alpro} & 19.16 & 33.58 & 77.27 & 25.0 & 54.55 & 24.8   \\
    \midrule
    \textbf{CEO-VQA} & 19.20 & 41.68 & 79.55 & 25.0 & 54.55 & 27.1 \\
    \bottomrule
    \end{tabular}
    }
    \label{tab:3}
    \vspace{-1em}
\end{table}



\section{Conclusion}
In this paper, we define a novel and challenging video question answering task, called Online Open-ended Video Question Answering (O$^2$VQA). To tackle this problem, we propose a Confidence-based Event-centric Online Video Question Answering (CEO-VQA) model. Specifically, a Confidence-guided Target Event Locator is designed to locate the target event and decide whether the collected information is enough to provide a confident answer. Then, a Target Event Question Answering model is developed to derive the final answer. To evaluate the proposed method, a large ATBS dataset is constructed. The experimental results on the ATBS dataset show that O$^2$VQA is a more challenging task and our method outperforms the state-of-the-art VideoQA model. 

\vfill
\clearpage

\small
\bibliographystyle{bib/IEEEbib}
\bibliography{bib/strings,bib/refs}

\end{document}

%% file: sections/code.tex
\begin{algorithm}[htpb]
  \SetKwFunction{Union}{Union}
  \SetKwFunction{VideoTextEncoder}{VideoTextEncoder}
  \SetKwFunction{VideoEncoder}{VideoEncoder}
  \SetKwFunction{QuestionEncoder}{QuestionEncoder}
  \SetKwFunction{MultiModalEncoder}{MultiModalEncoder}
  \SetKwFunction{AnswerDecoder}{AnswerDecoder}
  \SetKwFunction{Concat}{Concat}

  \SetKwInOut{Input}{Inputs}\SetKwInOut{Output}{Output}

  \Input{Online video stream $\mathcal{O}$, question $\bm{q}$, answer space $\mathcal{A}$}
  \Output{Answer $\bm{\alpha}$ to the question}
  Initialize confidence thresholds $c_{min}$, $c_{max}$\;
  \For{Every frame $\bm{f}\in \mathcal{O}$}{

    $c\leftarrow$ \VideoTextEncoder{$\bm{q}, \bm{f}$}\;
    \If{$c \ge c_{max}$}{
        $\mathcal{T}_{B} := $ \emph{Backward-traverse frames till $c < c_{min}$}\;
        $\mathcal{T}_{F} := $ \emph{Forward-traverse frames till $c < c_{min}$}\;
        $\mathcal{T} := $ {$\mathcal{T}_{B} \cup \mathcal{T}_{F}$}\;
        Break;
        }
  }
  
  \emph{Video features} $\bm{V} :=$\VideoEncoder{$\mathcal{T}$}\;
  \emph{Question features} $\bm{Q} :=$\QuestionEncoder{$\bm{q}$}\;
  $\bm{M} := $\MultiModalEncoder{$\bm{Q},\bm{V}$}\;
  $\bm{\alpha}\leftarrow$\AnswerDecoder{$\bm{M}$, $\mathcal{A}$}\;
  
  \caption{Confidence-based Event-centric Online Video Question Answering.}
  \label{algo:1}
\end{algorithm}
\vspace{-1.5em}